\documentstyle[aps,epsf,twocolumn]{revtex}
\renewcommand{\vec}[1]{\mbox{\boldmath$#1$}}
\begin{document}
\twocolumn[\hsize\textwidth\columnwidth\hsize\csname
@twocolumnfalse\endcsname
\title{
Spin-Blockade in Single and Double Quantum Dots in Magnetic Fields:\\
a Correlation Effect}\par

\author{Hiroshi Imamura, Hideo Aoki, and Peter A. Maksym$^{(a)}$}
\address{Department of Physics, University of Tokyo, 
Hongo, Tokyo 113, Japan \\ 
$^{(a)}$Department of Physics and Astronomy,
University of Leicester, Leicester LE1 7RH, United Kingdom}

\maketitle

\begin{abstract}
The total spin of correlated electrons in a quantum dot 
changes with magnetic field and this effect is
generally linked to the change in the total angular momentum from 
one magic number to another, which 
can be understood in terms of 
an `electron molecule' picture for strong fields. 
Here we propose to exploit this fact 
to realize a spin blockade, i.e., 
electrons are prohibited to tunnel at specific values of the magnetic field.  
The spin-blockade regions have been obtained by calculating both the 
ground and excited states. In double dots the spin-blockade condition 
is found to be less stringent than in single dots.
\end{abstract}


\vskip2pc]

The Coulomb blockade is one of the highlights 
in the transport properties of mesoscopic systems such as quantum dots. 
This is a combined effect of the discreteness of energy levels 
and the electron-electron interaction (charging energy).  
Now, it has recently been suggested that, 
if the total spins of the ground state of $N$ and $(N-1)$-electrons 
differ by more than 1/2, the dot is blocked with the corresponding peak 
in the conductance missing at zero temperature. This is 
called the spin-blockade~\cite{weinmann,tanaka} and has been studied 
theoretically for weak electron interaction regimes.  
There the Hund's coupling picture, in which electrons are 
accommodated in one-electron states with high spins for degenerate 
states, tends not to realize the spin-blockade condition, 
so that some modifications such as 
an anharmonicity in the confinement
potential~\cite{eto} have to be introduced.  

When quantum dots are placed in strong magnetic
fields, the ground states are known to 
change dramatically into the magic-number
states~\cite{girvin,map_prl}.  This comes from 
the electron correlation effect, 
since the magic numbers for the total angular momentum 
arise from a combined effect of the 
electron correlation and Pauli's principle, persisting 
even when the Zeeman energy is completely ignored.  
The total angular momentum of the ground state jumps from one
magic number to another as the magnetic field $B$ is varied.  

\begin{figure}[h]
    \epsfxsize=\columnwidth
\centerline{\hbox{
      \epsffile{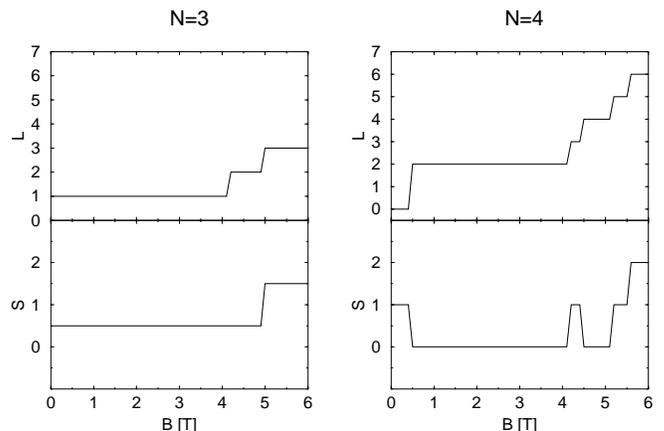 }  }}
  \caption{The total angular momentum and total spin of 
    the ground state for a single quantum dot with $N=3$ 
    (left) or $N=4$ electrons (right).
    The confinement potential is assumed to be parabolic with
    $\hbar\omega_{0} = 6.0$ meV.
}
  \label{fig:gs_gl_n3_n4_single}
\end{figure}

An important hint that electron correlation is really at work 
is the fact that the total spin ($S$), where $\vec{S}^{2}=S(S+1)$, of the
ground state, which dominates how the electrons correlate, 
changes wildly as shown in Fig.~\ref{fig:gs_gl_n3_n4_single}.  
This happens when the typical Coulomb energy is much greater than 
the single-electron level spacing, where electron 
molecule are formed.  In this sense this is genuinely 
an electron-correlation effect 
--- electron correlation 
has been known to dominate the spin states in ordinary correlated 
electron systems such as the Hubbard model, but the present case 
is a peculiar manifestation in strong magnetic fields.

In the present paper we propose to utilize this electron correlation effect to 
realize a spin blockade.  
We have numerically studied the ground and excited states of single
dots that contain three or four electrons with a
parabolic confinement potential and
find that the spin blockade should indeed be observed.  
Physically, a key observation starts from the fact 
that the correlated electron states 
in the dot may be thought of as `electron molecules'\cite{map_eckart}, 
which in turn enables us to interpret\cite{dresden} the spin wavefunctions 
taking part in the spin blockade as spin configurations in molecules, 
which include the resonating valence bond (RVB) states, 
that are usually invoked for lattice fermions.  
We further show that 
the spin-blockade condition is easier to satisfy in double dots
which can be tuned by 
controlling the layer separation and the strength of 
the interlayer tunneling.  

So let us start with looking at the total angular momentum ($L$) of 2D
electrons 
confined in a quantum dot in a magnetic field , which
has a one-to-one correspondence with the spatial extent ($\propto 
\sqrt{L}$) of the wavefunction.  Thus the presence of magic $L$ values
signifies that the total Coulombic energy of the interacting electrons, 
although roughly a decreasing function of $L$ 
as the electrons move further apart for larger $L$, 
is not a smooth 
function of the size of the wavefunction, 
so that jumps in $L$ are accompanied by 
jumps in the size of the wave function \cite{galejs,hallam}.  
For example, the total
angular momentum $L$ of three spin-polarized electrons changes $3
\rightarrow 6 \rightarrow 9 \rightarrow \dots$ 
with increasing magnetic field.  

Recently one of the authors has explained this as 
an effect of correlation in the electron configuration, 
where Pauli's 
exclusion principle dictates group-theoretically the manner in which the 
quantum numbers should appear~\cite{map_eckart}.  
There, the picture of the `electron molecule', in which the electrons 
with a specific configuration (triangle for three electrons, 
square for four, etc) rotating as a whole has turned out to 
be surprisingly accurate.  
This continues to be the case for larger numbers of 
electrons~\cite{hima3}. 

When one considers the spin degrees of freedom, 
the magic $L$ values are linked with 
the total spin. This is already apparent in the 
first numerical study of spin dependent correlation in quantum dots
~\cite{map_prb}.
These molecules are characterized by a quantum number, $k_s$, 
where the spin wave function $\Psi_{\rm spin}$ is transformed to 
${\rm exp}(-2\pi k_s i/m) \Psi_{\rm spin}$ under the rotation of 
$2\pi/m$ for an $m-$fold symmetric molecule.  
Then the criterion for the magic number, modulo $m$, reads
$L + k_s \equiv 0 (m/2)$ for $m$ odd (even).

To actually obtain  the spin states numerically for 
different numbers of electrons, 
let us consider single GaAs quantum dots 
with three or four electrons in a parabolic potential.  
The electron motion is assumed to be completely two dimensional.  
The Hamiltonian for a single dot is
${\cal H}={\cal H}_{\rm s}+{\cal H}_{\rm C}$, 
where 
\begin{equation}
  {\cal H}_{\rm s} = \sum_{n}\sum_{\ell}\sum_{\sigma}
  \varepsilon_{n \ell \sigma}
  c_{n \ell \sigma}^{\dag}c_{n \ell \sigma},
\end{equation}
is the single-electron part, while 
\begin{eqnarray}
  {\cal H}_{\rm C} &=&
  \frac{1}{2}\sum_{n_{1} \sim n_{4}}\sum_{\ell_{1}\sim\ell_{4}}
  \sum_{\sigma_{1}\sim\sigma_{4}}\nonumber\\
 &&\mbox{\ }
 \langle n_{1} \ell_{1}\sigma_{1},n_{2} \ell_{2}\sigma_{2} |
  \frac{e^{2}}
  {\epsilon |{\bf r}_{1}-{\bf r}_{2}|}
  |n_{3} \ell_{3}\sigma_{3},n_{4}\ell_{4}\sigma_{4}\rangle\nonumber\\
  &&\times  c_{n_{1}\ell_{1}\sigma_{1}}^{\dag}
  c_{n_{2}\ell_{2}\sigma_{2}}^{\dag}
  c_{n_{4}\ell_{4}\sigma_{4}} c_{n_{3}\ell_{3}\sigma_{3}}.
\end{eqnarray}
represents the Coulomb interaction.  
Here the Hamiltonian is written in 
second quantized form in a Fock-Darwin~\cite{fock} basis, 
and $\varepsilon_{n\ell\sigma}$$=$$(2n+1+|\ell|)$$\hbar$
$(\omega_{c}^{2}/4+\omega_{0}^{2})^{1/2}$$-\ell\hbar\omega_{c}/2$$
- g^{*}\mu_{_{B}} B s_{z}$. The dielectric constant is $\epsilon$,
$\hbar\omega_{0}$ represents the strength of the parabolic
confinement potential, $\omega_{c}=eB/m^{*}c$ is the cyclotron
frequency, $m^{*}$ is the effective mass, $\mu_{_{B}}$ is the Bohr
magneton, $g^{*}$ is the effective $g$-factor and $s_{z}$ is
$z$-component of the spin of a single electron.

We use the confinement potential $\hbar\omega_{0}=6.0$ meV. This is a 
little larger than usually estimated values $(2\sim4$ meV) and 
is deliberately chosen to reproduce the addition energy
spectrum~\cite{tarucha}. 
The fact that calculations with a $1/r$ interaction
require a larger confinement energy to reproduce 
experimental results is considered to be a consequence of
the modification of the interaction potential in real dots
\cite{Maksym97}. 

In our numerical calculations
we have used enough states (including 
higher Landau levels) in the basis to ensure 
convergence of the ground-state energy within $0.1\%$.  
Three lowest excited states are also calculated 
for each value of $B$, which turn out to be the lowest-energy states having 
different angular momenta in the present case.
Excited states are also obtained with a typical accuracy of 
$<0.1\%$ for an $N=3$ single dot at $B=5$ T.

The total angular momentum, and the total spin of the ground state 
for three- and four-electron systems plotted in
Fig.~\ref{fig:gs_gl_n3_n4_single}, we can 
see how the magic $L$ values go hand in hand with $S(N)$ for $N$
electrons, where $\vec{S}^{2}=S(S+1)$ while the $z$ component of
$\vec{S}$ is aligned to $\vec{B}$: 
As the magnetic field increase, 
the ground state changes as 
$(L,S)=(1,1/2)\rightarrow(2,1/2)\rightarrow(3,3/2)$ for $N=3$, 
$(L,S)=(0,1)\rightarrow(2,0)\rightarrow(3,1)\rightarrow(4,0)\rightarrow(5,1)\rightarrow(6,2)$ for $N=4$.  

\begin{figure}[h]
    \epsfxsize=0.8\columnwidth
\centerline{\hbox{
      \epsffile{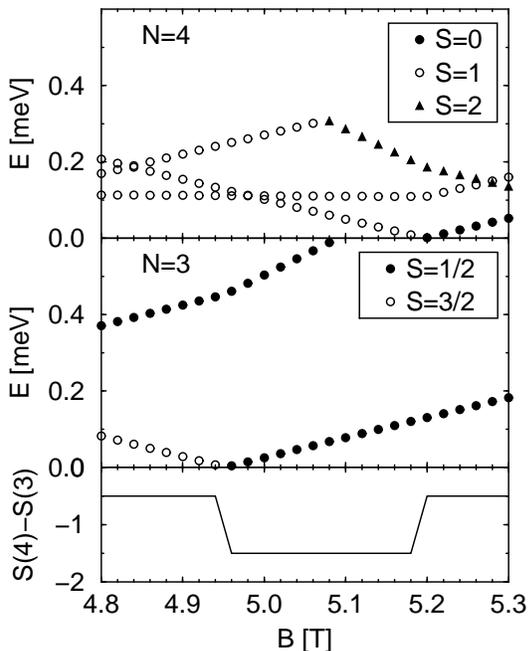}  }}
  \caption{
    Top: Excitation energies for $N=4$ single dot.
    Middle: The same for $N=3$. 
    Bottom: Difference in the total spin, $S(4)-S(3)$, 
    between the ground state for
    $N=3,4)$.
    }
  \label{fig:gs_exc_single}
\end{figure}

If we then plot the 
difference in the total spin, $S(4)-S(3)$, against the magnetic field
in the bottom panel of Fig.~\ref{fig:gs_exc_single}, the 
spin blockade condition,
\begin{equation}
|S(N)-S(N-1)|>\frac{1}{2},
\label{eq:spin_blockade}
\end{equation}
is indeed fulfilled: $S$ jumps from 3/2 to 0 
in the region $4.96 <B< 5.18$ T.

From the magic-number criterion 
the state with $(L=2+4\times{\rm integer})$ has to have 
the quantum number $k_s=0$, 
while the $(L = 4\times{\rm integer})$ state $k_s=2$ for $N=4$.
We can make an 
intriguing identification, 
by looking at the spin density correlation function, 
that $(L,S)=(2,0)$ is an ${\rm RVB}^{-}$ state while 
$(4,0)$ is an ${\rm RVB}^{+}$, where 
the RVB's are defined, for a four-site cluster, as
\begin{eqnarray}
\mbox{\raisebox{0.5\baselineskip}{RVB$^{-}$:}}&&\mbox{ 
 \epsfysize=1.5\baselineskip
   \epsffile{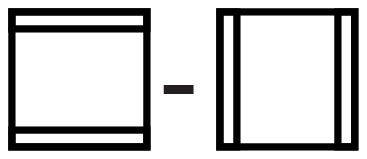}
}\nonumber\\
&&=|^{\uparrow \uparrow}_{\downarrow \downarrow} \rangle
       \! + \! |^{\downarrow \downarrow}_{\uparrow \uparrow} \rangle
       \! + \! |^{\uparrow \downarrow}_{\uparrow \downarrow} \rangle
       \! + \! |^{\downarrow \uparrow}_{\downarrow \uparrow} \rangle
      \! -2\! \left[ |^{\downarrow \uparrow}_{\uparrow \downarrow} \rangle
       \!+ \!|^{\uparrow \downarrow}_{\downarrow \uparrow} \rangle \right],
\\
\mbox{\raisebox{0.5\baselineskip}{RVB$^{+}$:}}&&\mbox{ 
 \epsfysize=1.5\baselineskip
   \epsffile{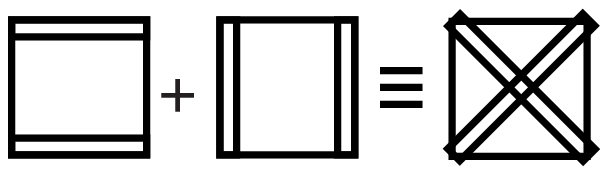}
}\nonumber\\
&&= |^{\uparrow \uparrow}_{\downarrow \downarrow} \rangle
        + |^{\downarrow \downarrow}_{\uparrow \uparrow} \rangle
        - |^{\uparrow \downarrow}_{\uparrow \downarrow} \rangle
        - |^{\downarrow \uparrow}_{\downarrow \uparrow} \rangle,
\end{eqnarray}
with 
\raisebox{-0.1\baselineskip}{
\mbox{
 \epsfysize=0.5\baselineskip
   \epsffile{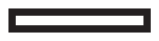}
}}
$\equiv
  \frac{1}{\sqrt{2}}(|\uparrow\downarrow \rangle - |\downarrow\uparrow
  \rangle )$ being the spin-singlet pair 
in the electron molecule.  
The difference of RVB$^+$ from RVB$^-$ is that the former 
lacks the N\'{e}el components (the last two terms in RVB$^-$) and has
the extra phase factor -1 for $\pi/2$ rotation.  
Although 
what we have here is totally different from 
lattice fermion systems such as the Hubbard model 
for which RVB is usually conceived, 
the electron-molecule formation has brought about such spin configurations.

In this region, 
the conduction, blocked at zero temperature, has to occur through 
an $S=1$ excited state for $N=4$  at finite temperatures.  
If the excited states are well separated in energy 
($\geq$ 0.1 meV, typical experimental resolution for Coulomb diamonds) 
from the ground state 
for both of the $(N-1)-$ electron and $N-$ electron states,
the spin blockade should be observed in the Coulomb diamond, which is the 
differential conductance plotted in the plane of source-drain voltage 
and gate voltage.  
We have calculated the three lowest excitation energies and their total
spins for the $N=4$ quantum dot in Fig.~\ref{fig:gs_exc_single}.
The lowest excited states for $N=3$ and for $N=4$ both lie 
about 0.06meV above the ground state around $B=5.1$ T in the
spin-blockade region.  
We can make this separation larger ($\sim$ 0.1
meV) for stronger confinement potentials (e.g., 0.09meV 
around $B=7.4$ T for $\hbar\omega_{0}=8.0$ meV). 
Such confinement potentials may be realized 
in a gated vertical quantum dot~\cite{tarucha}.

The link between the magic $L$ and total $S$ and subsequent 
spin blockade appears for other numbers of electrons as well, 
e.g., between  $(L,S)=(2,0)$ state
for $N=2$ and $(6,3/2)$ state for $N=3$ for $14.1<B<14.8$ T.

Now we move on to the double dots, where 
dots are separated in the vertical direction with their centers
aligned on a common axis.  
We assume the same confinement potential for the two dots for simplicity. 
Here electrons are Coulomb-correlated both within each layer and across 
the two layers, in the presence of the inter-layer tunneling.  
Recent advances in semiconductor fabrication techniques have 
enabled fabrication of double dots in vertical, triple-barrier 
structures on submicron scales~\cite{austing}.  
Theory for the double quantum dots has 
been developed~\cite{hima1,hima2,oh,palacios,dagotto,tamura}, 
where intriguing features such as magic-number 
states intrinsic to double dots, or a singlet-to-triplet 
spin transition for two-electron system have been 
found~\cite{hima1,hima2,oh}.

The Hamiltonian now contains the tunneling term, 
\begin{equation}
  {\cal H}_{\rm t} = -\frac{\Delta_{\rm SAS}}{2}\sum_{n}\sum_{\ell}
  \left(
    c_{n \ell +}^{\dag}c_{n \ell -} + c_{n \ell -}^{\dag}c_{n \ell +}
  \right) ,
\end{equation}
while the Coulomb part is now the matrix element of 
$e^2/\epsilon |{\bf r}_{1}-{\bf r}_{2}|$ 
for intra-layer interaction, and
$e^2/\epsilon (|{\bf r}_{1}-{\bf r}_{2}|^2 + d^2 )^{1/2}$ 
for inter-layer interaction. The basis is
$| n_{i} \ell_{i}\sigma_{i}\alpha_{i}\rangle$,
where $\alpha= \pm$ is an index specifying the two dots.

\begin{figure}[h]q
    \epsfxsize=\columnwidth
\centerline{\hbox{
      \epsffile{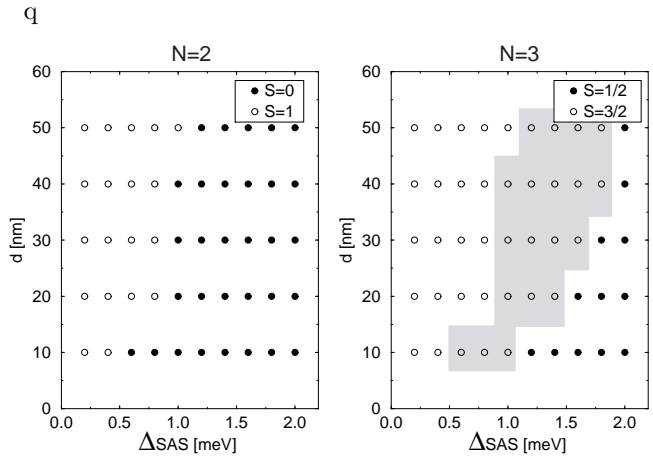}  }}
  \caption{
   The ground-state spin of the double dot for $N=2$(left) or $N=3$(right) 
   for $B=5$ T
   for a parabolic confinement potential $\hbar\omega_{0} = 6.0$ meV. 
   Shaded region corresponds to a transition from $S=0 \rightarrow 3/2$
 }
  \label{fig:gs_phase_double}
\end{figure}

Thus a double dot is characterized by the
parabolic confinement potential layer, 
$\hbar\omega_{0}$, the layer separation, $d$, and
the strength of the inter-layer tunneling (measured by $\Delta_{\rm SAS}$,
the energy gap between the symmetric and antisymmetric one-electron states).  
Here we have adopted realistic values of $\hbar\omega_{0}=6.0$
meV, 10 $\leq d \leq$ 50 nm, 0.2 $\leq \Delta_{\rm SAS}\leq$ 
2.0 meV.  
We can now plot in 
Fig.~\ref{fig:gs_phase_double}
how high-spin states appear on the $\Delta_{\rm SAS}-d$ plane.  
A high-spin state is indeed seen to appear in the upper left region of 
each panel for $B=5.0$  T.  
In the shaded region of the right
panel, the difference between the total spins is $S(3)-S(2)=3/2$, 
fulfilling the spin blockade condition.

\begin{figure}[h]
    \epsfxsize=\columnwidth
\centerline{\hbox{
      \epsffile{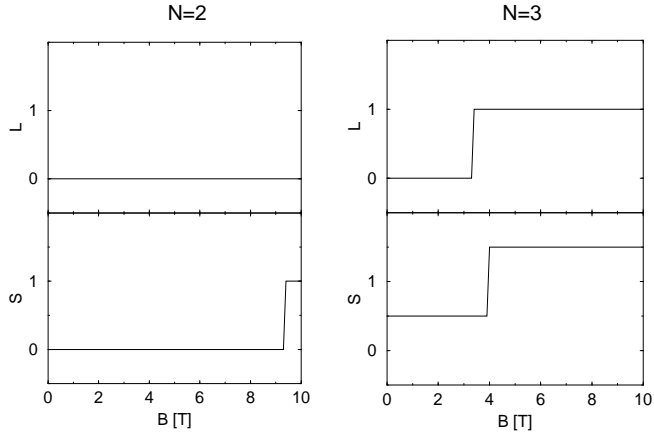}  }}
  \caption{The total angular momentum and total spin of 
    the ground state for double dots with $N=2$(left) $N=3$(right) 
    electrons.  
    $\hbar\omega_{0} = 6.0$ meV. $d = 16.0$ nm, and
    $\Delta_{\rm SAS}=1.2$ meV.
    }
  \label{fig:gs_gl_n2_n3_double}
\end{figure}

We now focus on a typical point in the shaded region, 
$d = 16.0$ nm and $\Delta_{\rm SAS}=
1.2$ meV.  In Fig.~\ref{fig:gs_gl_n2_n3_double} the total energy, 
total angular momentum and 
the total spin of the ground state for $N=2, 3$ are plotted.

\begin{figure}[h]
    \epsfxsize=0.8\columnwidth
\centerline{\hbox{
      \epsffile{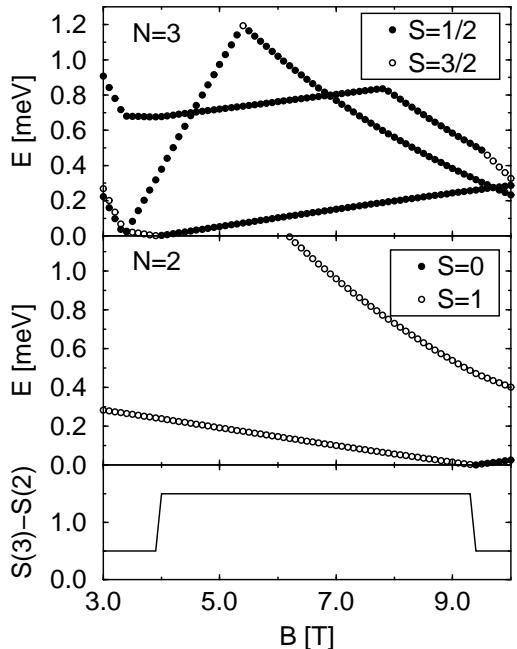}  }}
  \caption{Top (middle): Excitation energies for $N=3$\newline$ (N=2)$ double
    dots.  Bottom: The difference, $S(3)-S(2)$, in the total spin for 
    $N=2$ and $N=3$ double dots.
    }
  \label{fig:gs_exc_double}
\end{figure}

The difference between the total spin of two and three electrons
systems is shown in the bottom panel of Fig.~\ref{fig:gs_exc_double}.
The spin-blockade condition is satisfied for 4.0 $\leq B \leq 9.3$T, 
which is wider than for the single dot.  
In the bulk bilayer fractional quantum Hall (QH) systems, 
a phase diagram on the 
$\Delta_{\rm SAS}-d$ plane has been considered.  
If we translate\cite{dresden} the quantities 
for the dots, we are working in the `two-component' (correlation-dominated) 
region around the QH-non QH boundary in the language for the 
bilayer QH system.  
This might have some relevance to the behavior of the double dots.  
In Fig.~\ref{fig:gs_exc_double}, the excitation 
energies for the $S=1/2, 3/2$ states are also plotted.  The excitation energies
for both $N=2$ and $N=3$ systems are about 0.12 meV (exceed 0.1 meV) at
$B=6.4$ T, which
is large enough for the spin blockade to be observed.  
We also notice a level crossing between the second and the third 
excited states around $B=6.9$ T for $N=3$, which should appear 
in the Coulomb diamond.

In summary, we have shown that in both 
single and double dots, a spin blockade should occur 
in some magnetic field region, as an effect of the 
total spin dominated by the magic angular momenta.  
We would like to thank to Seigo Tarucha and Guy Austing for
a number of valuable discussions.


\end{document}